\documentclass{pnastwo}

\usepackage[dvips]{graphicx}
\usepackage{amssymb,amsfonts,amsmath,color}
\usepackage{color}
\usepackage{hyperref}

\begin{document}

\title{Ciliary contact interactions dominate\\ surface scattering of swimming eukaryotes}

\author{
Vasily Kantsler\affil{1}{Department of Applied Mathematics and Theoretical Physics, Centre for Mathematical Sciences, University of Cambridge, Wilberforce Road, Cambridge CB3 0WA, UK},
J\"orn Dunkel\affil{1}{Department of Applied Mathematics and Theoretical Physics, Centre for Mathematical Sciences, University of Cambridge, Wilberforce Road, Cambridge CB3 0WA, UK},
Marco Polin\affil{1}{Department of Applied Mathematics and Theoretical Physics, Centre for Mathematical Sciences, University of Cambridge, Wilberforce Road, Cambridge CB3 0WA, UK},  
Raymond E. Goldstein\affil{1}{Department of Applied Mathematics and Theoretical Physics, Centre for Mathematical Sciences, University of Cambridge, Wilberforce Road, Cambridge CB3 0WA, UK} 
}

\contributor{Preprint as accepted for publication in PNAS, for published journal version (open access)  and Supporting Information  \href{http://dx.doi.org/10.1073/pnas.1210548110}{\textcolor{blue}{click here}}
}

\maketitle

\begin{article}

\begin{abstract}
Interactions between swimming cells and surfaces are essential to many microbiological processes, from bacterial biofilm formation to human fertilization. However, in spite of their fundamental importance,  relatively little is known about the physical mechanisms that govern the scattering of flagellated or ciliated cells from solid surfaces.  A more detailed understanding of these interactions promises not only new biological insights into structure and dynamics of flagella and cilia, but may also lead to new microfluidic techniques for controlling cell motility and microbial locomotion, with potential applications ranging from diagnostic tools to therapeutic protein synthesis and  photosynthetic biofuel production. Due to fundamental differences in physiology and swimming strategies, it is an open question whether microfluidic transport and rectification schemes that have recently been demonstrated for pusher-type microswimmers such as bacteria and sperm cells, can be transferred to puller-type algae and other motile eukaryotes, as it is not known whether long-range hydrodynamic or short-range mechanical forces dominate the surface interactions of these microorganisms. Here, using high-speed microscopic imaging, we present direct experimental evidence that the surface scattering of both mammalian sperm cells and  unicellular green algae is primarily  governed by direct ciliary contact interactions.  
 Building on this insight, we predict and verify experimentally the existence of optimal microfluidic ratchets that maximize rectification of initially uniform  \textit{Chlamydomonas reinhardtii} suspensions. Since mechano-elastic properties of cilia are conserved across eukaryotic species, we expect that  our results apply to a wide range of swimming microorganisms.
\end{abstract}



\dropcap{S}urface interactions of motile cells play crucial roles in a wide range of microbiological phenomena, perhaps most prominently in the formation of biofilms~\cite{1998Kolter} and during the fertilization of mammalian ova~\cite{2003Quill}. Yet, in spite of their widely recognized importance, the basic physical mechanisms that govern the response  of swimming bacteria,  algae or spermatozoa to solid surfaces have remained unclear. This predicament is  exemplified by the current debate~\cite{2008Berke_PRL,2009Tang_PRL,2011Tang_PRE,2011Drescher_PNAS} about the relevance of hydrodynamic long-range forces and steric short-range interactions for the accumulation of flagellated cells at liquid-solid interfaces. From a general perspective, improving our understanding of cell-surface scattering processes  promises not only new  insights into structure, dynamics and biological functions of flagella and cilia -- it will also help to advance microfluidic techniques for controlling microbial locomotion~\cite{2007AustinChaikin,2010Austin_PRL}, with potential applications in diagnostics~\cite{2012Denissenko},  therapeutic protein synthesis~\cite{2010Mayfield} and  photosynthetic biofuel production~\cite{2008Godman,2009Happe_PNAS,2011Melis,2011MayfieldWong}.  That microfluidic circuits provide an excellent testbed for developing and assessing new strategies for the control of cell motility was recently demonstrated by the rectification of random bacterial swimming through microscopic wedge-shaped barriers~\cite{2007AustinChaikin,2010Austin_PRL}. However, since eukaryotic and prokaryotic swimming strategies differ substantially from each other~\cite{2011Drescher_PNAS,2009Polin_Science,2010Drescher_PRL,2010Guasto_PRL, 2006Berg_Nature,Berg}, it is unclear whether design principles that exploit surface collisions to achieve control of bacterial locomotion are transferrable to motile eukaryotes.

\par
\begin{figure*}[b]
\centering
\includegraphics[clip=,width= 18cm]{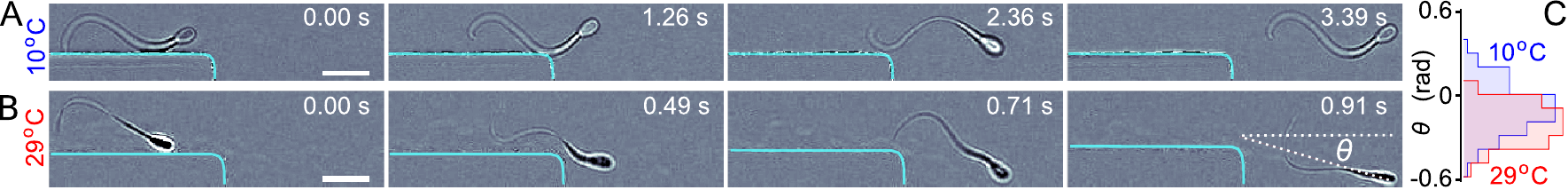} 
\caption{
Surface scattering of bull spermatozoa is governed by ciliary contact interactions, as evident from the scattering sequences of individual cells at two temperature values (A) $T=10^\circ$C and (B)  $T=29^\circ$C.  The background has been subtracted from the micrographs to enhance the visibility of the cilia. The  cyan-colored line indicates the corner-shaped boundary of the microfluidic channels (see Movies~S1 and S2 for raw imaging data).  The  horizontal dotted line in the last image in (B) defines $\theta=0$.  Scale bars $20\,\mu$m. (C)~The probability distributions of scattering angles $\theta$  from the corner peak at negative angles, due to the fact that the beat amplitude of the cilia exceeds the size of the cell body (sample size: $n=116$ for $T=10^\circ$C and  $n=115$ for $T=29^\circ$C).  At higher temperatures, the cilia exhibit a larger oscillation amplitude and beat frequency~\cite{1984Rikmenspoel}, resulting in a larger swimming speed and shifting the typical scattering angles to larger absolute values.\label{f:sperm_scattering}}
\end{figure*}

 
Aiming to elucidate the role of eukaryotic cilia in cell-surface interactions, we report here a detailed experimental investigation of surface scattering for bull spermatozoa and~\textit{Chlamydomonas reinhardtii} algae (simply referred to as {\it Chlamydomonas} in the following). Bull sperm and other mammalian spermatozoa  are \lq\lq pusher\rq\rq\space swimmers that generate propulsion by undulating a single posterior cilium (Fig.~\ref{f:sperm_scattering}A). By contrast, a wild-type {\it Chlamydomonas} cell is a \lq\lq puller\rq\rq\space that achieves locomotion by the breaststroke-like beating~\cite{2009Polin_Science,2010Drescher_PRL,2010Guasto_PRL} of a pair of anterior flagella (Fig.~\ref{f:chlamy_scattering}A).  {\it Chlamydomonas} algae have long been appreciated as premier model organisms in biology~\cite{1991Ruffer,2001Harris,2009Harris}, in particular for studying  photosynthesis~\cite{2008Godman,2007ChlamyGenome} and ciliary~\cite{2009Polin_Science,1991Ruffer,2011Fujiu} functions in eukaryotes. More recently, they have also attracted considerable interest as possible sources of  therapeutic proteins~\cite{2010Mayfield} and renewable biofuels~\cite{2008Godman,2009Happe_PNAS,2011Melis,2011MayfieldWong,2001MelisHappe,2006Esper,2010Posewitz_EukCell,2011Das}. Against this backdrop, our second goal  is to demonstrate the feasibility of microfluidic rectification schemes for these organisms. 
 
 \par 
Rectification of bacterial run-and-tumble motion in microfluidic ratchets~\cite{2007AustinChaikin,2010Austin_PRL} is believed to result from the swimmers' tendency to align their motion along the ratchet barriers, either by steric~\cite{2009Tang_PRL,2011Tang_PRE,2011Drescher_PNAS} or by hydrodynamic~\cite{2008Berke_PRL} surface  interactions, although the exact mechanism is not well understood.
While steric alignment with surfaces seems intuitively plausible for rod-shaped bacteria like {\it E. coli} or {\it B. subtilis}, additional hydrodynamic alignment is thought to arise from the fact that the posterior flagellar bundle of a bacterium creates a ``pusher''-like dipolar flow  during locomotion~\cite{Berg}. This flow points outwards along the body axis and inwards along the lateral directions~\cite{2011Drescher_PNAS}. 
The presence of a wall couples the swimmer's translation and rotation, and causes it to align parallel to the surface~\cite{2008Berke_PRL}. By contrast, the anterior flagella of wild-type {\it Chlamydomonas}  ``pull'' the organism through the fluid, thereby generating a far-field flow topology~\cite{2010Drescher_PRL,2010Guasto_PRL} that looks roughly opposite to that of a bacterium.
Hence, far-field hydrodynamics suggests that {\it Chlamydomonas} should either turn away from  or collide  head-on with a nearby no-slip surface, but the complex time-dependent flow structure~\cite{2010Drescher_PRL,2010Guasto_PRL} close to the cell body makes it difficult to predict the scattering dynamics in the vicinity of the surface. It is therefore not possible to infer from general hydrodynamic arguments whether it is at all feasible to design microfluidic structures that are capable of rectifying algal swimming.
Moreover, purely hydrodynamic considerations completely neglect  direct contact  interactions between cilia or flagella and solid surfaces. Unfortunately, this potentially important scattering mechanism~\cite{2009Tang_PRL} is not included in currently prevailing theoretical models of microbial swimming near solid boundaries~\cite{2008Berke_PRL}. 

\par
Here, we present  direct experimental evidence that the scattering of bull spermatozoa and  {\it Chlamydomonas} algae off a solid boundary is, in fact,  mainly determined by the contact interactions between their flagella and the surface, while hydrodynamic effects only play a secondary role. Building on these insights we derive a simple criterion to predict an efficient ratchet design for {\it Chlamydomonas} and confirm its validity experimentally, thereby demonstrating that robust rectification of algal locomotion is possible.  
More generally, our results show that the interactions between swimming microorganisms  and surfaces are more complex than previously recognized, suggesting the need for a thorough revision of currently accepted paradigms.  
Since mechano-elastic properties of eukaryotic cilia are conserved across eukaryotic species, we expect  flagella-surface interactions to play a similarly important role for a wide range of natural microswimmers, thus promising new diagnostic tools and microfluidic sorting devices for  sperm~\cite{2012Denissenko} and other motile cells.

\section{Results}

To identify the dynamical details of eukaryotic cell-surface interactions, we studied the surface scattering of bull spermatozoa and four different \textit{Chlamydomonas} algae strains in quasi-2D  microfluidic channels (height $\sim25\mu$m), using high-speed microscopic imaging (Materials and Methods).

\subsection{Scattering of Individual Sperm Cells from Solid Boundaries}
We analyzed the scattering of individual sperm cells from corner-shaped channel boundaries at high and low  temperature (Fig.~1A,B and Movies~S1 and S2). Direct observation reveals that the short-range interaction of the cilium with the boundary determines how a spermatozoon swims along a solid surface. In a typical scattering event,  a sperm cell closely follows the boundary until it reaches a corner and departs at an angle $\theta$, defined here relative to the initial swimming direction such that $\theta=0$ corresponds to the surface tangent (Fig.~1B). We determined the temperature-dependent distributions of $\theta$ from more than 200 scattering events by tracking the position of the cell body up to a distance of $70\,\mu$m from the corner (Fig.~1C). The histograms  show that sperm-surface interactions are typically characterized by negative scattering angles $\theta<0$, due to the fact that the beat amplitude of the ciliary motion is much larger than the size of the cell body. Hence, ciliary contact with the surface tends to turn the spermatozoa towards the boundary, thereby  preventing their escape from flat and weakly curved surfaces.
\par
To illustrate how differences in the ciliary beating patterns affect the surface interaction of spermatozoa, we exploit the fact~\cite{1984Rikmenspoel} that amplitude and frequency of the ciliary motion increase monotonously with temperature $T$ in the range $5^\circ\text{C}<T<38^\circ\text{C}$.  This variability is caused by a change in motor activity~\cite{2007RiedelKruse}, approximately described by an Arrhenius law $\propto \exp(-\Delta/kT)$~\cite{1984Rikmenspoel}, where $\Delta$ denotes the activation energy and $k$ the Boltzmann constant. If ciliary contact governs the surface interactions of sperm, then one should expect that the absolute mean scattering angle $|\bar{\theta}|$ increases with temperature.  By comparing the scattering distributions at low and high temperatures, we find that this is indeed the case (Fig. 1C). At low temperature  $T=(10\pm 1)^\circ$C  the experimental data yields a mean scattering angle $\bar{\theta}= (-5.6\pm 1.0)^\circ$, whereas $\bar{\theta}= (-12.6\pm 0.7)^\circ$ at high temperature $T=(29\pm 1)^\circ$C. In particular, these results suggest that, at higher temperatures, sperm cells can become more easily trapped at strongly curved surfaces. With regard to future biotechnological applications, this self-trapping by ciliary beating can provide a useful mechanistic basis for sorting and rectifying spermatozoa~\cite{2012Denissenko}. 

\begin{figure*}[b]
\centering 
\includegraphics[clip=,width= 18cm]{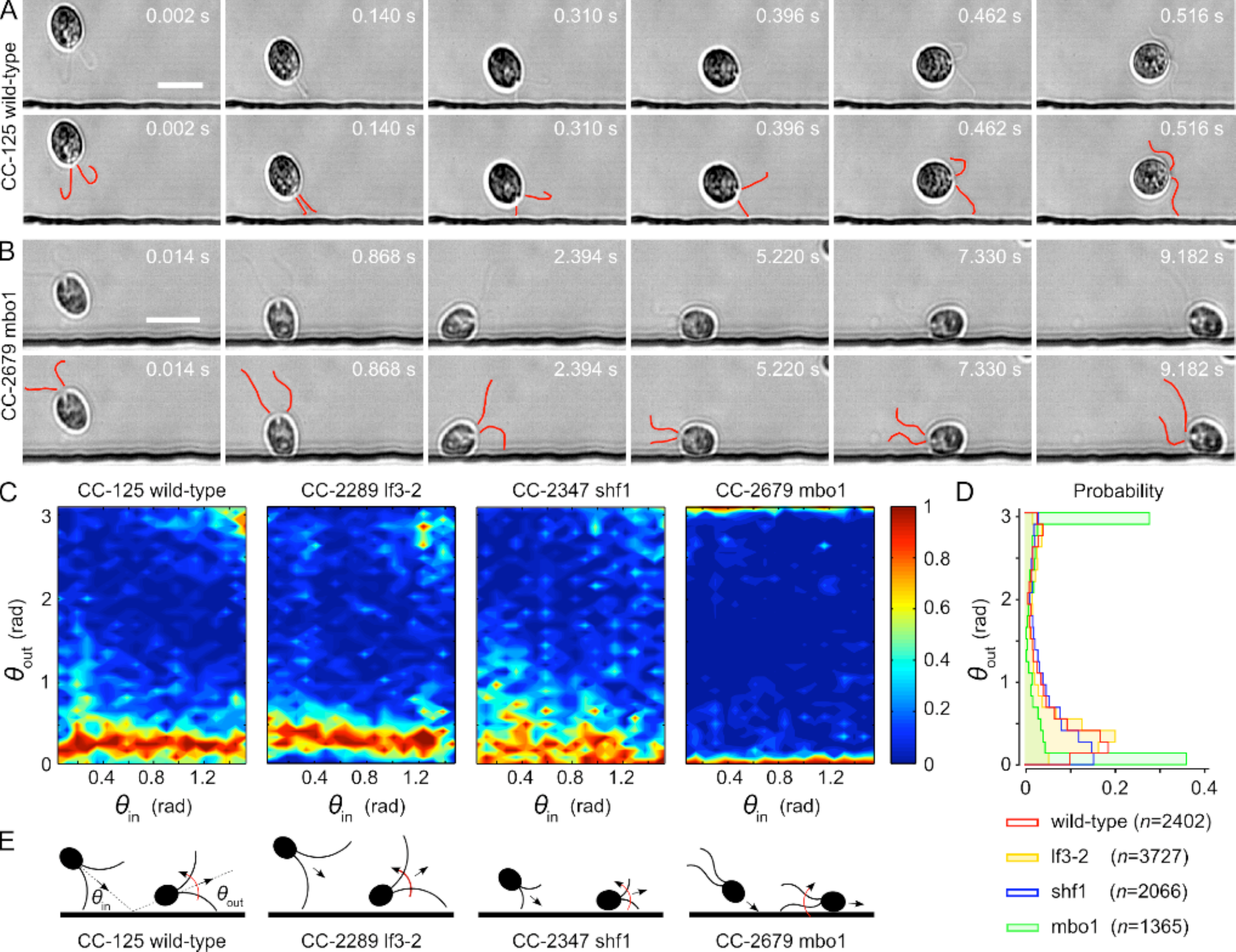} 
\caption{
Surface scattering of \textit{Chlamydomonas} is governed by ciliary contact interactions.
(A)~Scattering sequence for wild-type {\it Chlamydomonas} CC-125 (Movie~S3). Upper panel: original micrographs. Lower panel:  cilia manually marked red. Results for the long-flagella mutant lf3-2 and the short-flagella mutant shf1 look qualitatively similar (Movies~S4 and S5). Scale bar $20\,\mu$m.
(B)~The mutant pusher mbo1 remains trapped for several seconds (Movie~S6). Scale bar $20\,\mu$m.
(C)~The conditional probability distributions $P(\theta_\text{out}| \theta_\text{in})$  indicate that, for all four strains, memory of the incidence angle is lost during the collision process, due to multiple flagellar contact with the surface.
(D)~The cumulative scattering distribution $P(\theta_\text{out})$ shows how cilia length and swimming mechanisms determine the effective surface-scattering law.
(E)~Schematic illustration of the flagella-induced scattering and trapping mechanisms.
\label{f:chlamy_scattering}}
\end{figure*}
\subsection{Flagella-Induced Scattering of Individual Algae from Solid Boundaries}
To further test the idea that ciliary contact dominates eukaryotic cell-surface interactions, we studied the surface scattering of four different \textit{Chlamydomonas} strains (The Chlamydomonas Resource Center, \texttt{http://www.chlamy.org}): 
the wild-type  CC-125, the long-flagella mutant CC-2289 lf3-2, 
the short-flagella mutant CC-2347 shf1, and the moving-backwards-only mutant CC-2679 mbo1 (Fig.~\ref{f:chlamy_scattering}). All four \textit{Chlamydomonas} strains share an essentially identical geometrical structure, but while the wild-type and the mutants lf3-2 and shf1  are ``puller''-type swimmers that differ only in flagella length ($6-8\mu$m for shf1~\cite{1987Kuchka}, compared to $11-13\,\mu$m for wild-type and  $12-22\,\mu$m for  lf3-2~\cite{1988Barsel}), the backwards-swimmer mbo1 has 
a persistent undulatory swimming gait and can be considered a ``pusher''  similar to bacteria and spermatozoa (see Movies~S3, S4, S5, S6, S7).  For each strain, we recorded more than $1300$ boundary-scattering events in quasi-2D microfluidic channels (height $\sim25\mu$m). 

High-speed imaging of individual cell  trajectories for both wild-type and mutants reveals that the interaction of  \textit{Chlamydomonas} with the channel wall is also strongly affected by the short-range contact forces between the flagella and the surface  (Fig.~\ref{f:chlamy_scattering}A,B and Movies~S3, S4, S5, S6).  In the case of the three ``puller'' swimmers (wild-type, lf3-2 and shf1),  the flagella prevent the cell body from touching the surface while simultaneously creating an effective torque that turns the organism away from it (Fig.~\ref{f:chlamy_scattering}A,E).  By contrast, for the backwards-swimming mutant mbo1, posterior thrust by the flagella  pushes the cell body onto the surface (Fig.~\ref{f:chlamy_scattering}B,E). Subsequently, the mechanical contact of the cilia with the boundary leads to a net torque that keeps rotating the alga towards the surface (Fig.~\ref{f:chlamy_scattering}E). As a result, mbo1 cells remain trapped at the channel wall for several seconds, compared to $\lesssim0.5\,$s for wild-type. 

\par
The contact force that is exerted by a flagellum onto the surface per stroke can be estimated 
from the flagellar beat frequency ($\sim50$ Hz) and the experimentally observed angular displacement amplitude per beat ($\sim0.2$ rad), which gives a typical angular speed $\omega\sim10$~rad/s. Assuming a spherical cell body (radius $a\sim5\,\mu$m), the torque $T$  can be obtained from $T\sim \xi \omega$, where the rotational drag coefficient is given by $\xi=8\pi\eta a^3$. Using $\eta= 10^{-3}$ Pa$\cdot$s for water, we find $T\sim 30\,$pN$\cdot\mu$m and, furthermore, by assuming a flagella length $L\sim10\,\mu$m, the typical force $F\sim T/L\sim 3\,$pN. These estimates, which are based on the observed rotation of the cell body near the surface, are consistent with the values obtained from recent measurements of freely swimming  \textit{Chlamydomonas} algae~\cite{2010Guasto_PRL,2011Bayly}, suggesting that reorientation at the wall is primarily determined by flagellar contact.

\par 
To quantify the different surface scattering laws for  each of the four \textit{Chlamydomonas} strains in  detail, we measured the incidence and scattering angles $\theta_\text{in}$ and $\theta_\text{out}$ (see Materials and Methods) and determined the conditional scattering distributions $P(\theta_\text{out}| \theta_\text{in})$, defined as the probability of being scattered into the interval $[\theta_\text{out}, \theta_\text{out}+d\theta_\text{out}]$ for a given incidence angle~$\theta_\text{in}$  (Fig.~\ref{f:chlamy_scattering}C). For wild-type \textit{Chlamydomonas},  $P(\theta_\text{out}| \theta_\text{in})$ is independent of the incidence angle~$\theta_\text{in}$ and exhibits a narrow peak at $\overline{\theta}_\text{out}\sim16^\circ$. The typical scattering angle $\overline{\theta}_\text{out}\sim20^\circ$ is larger  for the long-flagella mutant lf3-2,  whereas for the short-flagella mutant shf1, the maximum of $P(\theta_\text{out}| \theta_\text{in})$  is shifted to smaller angles $\overline{\theta}_\text{out}\sim12^\circ$. The systematic increase of the typical scattering angle with flagella length, in conjunction with the above force estimates, implies that, in all three cases, the characteristic escape angle~$\overline{\theta}_\text{out}$ is set geometrically  through the length of the cilia, the diameter of the cell body  and the typical distance of the latter from the surface at the moment of departure (see Movies~S3, S4, S5). For shf1, the cell body can come closer to the wall than for the wild-type, and the resulting lubrication forces may also affect the escape dynamics. The relatively larger spread in the distribution of scattering angles for shf1 is compatible with previous observations of stronger intrinsic fluctuations in shorter flagella~\cite{goldstein11}. By contrast, the ``pusher''-mutant mbo1 generally remains close to the surface, $\theta_\text{out}\sim0$ or   $\theta_\text{out}\sim\pi$ (Movie~S6). The differences between the scattering laws of the four different strains are also clearly evident from the mean scattering distributions $P(\theta_\text{out})$, obtained by averaging $P(\theta_\text{out}| \theta_\text{in})$ over all incoming angles  (Fig.~\ref{f:chlamy_scattering}ED. The symmetric bimodal shape of $P(\theta_\text{out})$ for mbo1 signals a complete loss information about the incidence angle, whereas in the case of the three \lq\lq puller\rq\rq\space strains the swimmers still remember their incoming directions, but not the exact values of~$\theta_\text{in}$.  Quantitatively similar  probability distributions characterize scattering laws in thicker chambers ($80\,\mu$m and $300\,\mu$m).

\subsection{Optimal Rectification of Algal Locomotion in Microfluidic Ratchets}
Knowledge of flagella-induced scattering can be used to design microscopic obstacles that will passively guide the random swimming of microorganisms in a desired direction (Fig.~3).
The design principles for mbo1 mutants and spermatozoa~\cite{2012Denissenko} are similar to those for  \textit{E. coli}~\cite{2007AustinChaikin,2010Austin_PRL}, as these species glide along surfaces after collision.
The challenge is to find the optimal rectification geometry for ``puller'' organisms, which scatter at a finite angle $\theta_\text{out}>0$ off solid surfaces.
To demonstrate the optimization procedure, we focus on the wild-type  {\it Chlamydomonas} strain, which has the peak angle $\overline{\theta}_\text{out}\sim 16^\circ$, and consider wedge-shaped obstacles, like those successfully employed in bacterial rectification~\cite{2007AustinChaikin,2010Austin_PRL} (see Fig.~3B). Optimal rectification requires maximizing the ratio~$j_{F}/j_{B}$ between the algae currents in forward and backward directions. The diffusive  backward flux $j_B$ is driven by gradients in the algal concentrations across the obstacle rows  (Fig.~3C)  and can be minimized by decreasing the gap distance $d_G$ between neighboring obstacles relative to the effective width $d_B$ of a single barrier,  $d_G\ll d_B$ (Fig. 3B). At the same time, the forward current $j_F$ can be maximized by adjusting the wedge angle $\alpha$ of the barriers to exploit secondary scattering (Fig. 3B).  Assuming a fixed angle ${\theta}_\text{out}$, basic geometric considerations yield the criterion $2 \alpha + {\theta}_\text{out} \leq \pi/2$ for secondary scattering in the forward direction. Thus,  for ideal deterministic scatterers with ${\theta}_\text{out}=\overline{\theta}_\text{out}$ one would expect maximal rectification for  $\alpha_*\approx \pi/4-\overline{\theta}_\text{out}/2$, yielding $\alpha_*\approx  37^\circ$ for wild-type {\it Chlamydomonas}. This estimate should, however,  be viewed as an upper bound for the optimal opening angle, since the measured scattering distributions  $P(\theta_\text{out})$ have large-angle tails  (Fig.~\ref{f:chlamy_scattering}D) that shift the optimal  $\alpha$-value to smaller angles.

We tested this prediction in numerical simulations of a minimal 2D model and by performing quasi-2D microfluidic experiments (Fig.~3). In our simulations (see Materials and Methods for details), microswimmers are represented  by non-interacting point particles whose motion captures the main observed features of {\it Chlamydomonas} trajectories.  Each particle  moves ballistically at a constant speed $V$,  performs random turns after exponentially distributed run-periods~\cite{2009Polin_Science} with persistence time $\tau\sim 1.5\,$s, and scatters from boundaries at a random angle ${\theta}_\text{out}$. Both  $V$ and ${\theta}_\text{out}$ are drawn from distributions that mimic the experimentally measured speed distributions and scattering laws (see Materials and Methods). The value of $\tau$ is consistent with estimates from the  mean square displacement of wild-type {\it Chlamydomonas}, as measured in the quasi-2D microfluidic chambers (height $25\mu$m) used for the rectification experiments (Fig.~3A).  Compared with 3D chambers,  the typical swimming speed of the algae ($\bar{V}\sim 30\mu$m/s in quasi-2D chambers) is reduced by approximately a factor of two due to the presence of nearby no-slip surfaces, which also tend to suppress hydrodynamic interactions~\cite{1976Liron}.

In both experiments and simulations, we considered chambers with four compartments separated by  rows  of wedge-shaped barriers with gaps of length $d_G$ (Fig.~3B,C).
For each wedge angle $\alpha$, the arm length $L$ of the barriers was chosen such that the length of the longest possible deterministic trajectory leading through the gap after scattering is equal to the persistence path-length $L_p$ of the algae (Fig.~3B). We then systematically quantified the rectification efficiency by scanning the $(d_G,\alpha)$ parameter space (Fig.~3D). Here, the rectification efficiency is defined by the ratio $R=\langle N_4 \rangle/\langle N_1 \rangle$, where  $\langle N_{i} \rangle$  denotes the steady-state time-average of the number of swimmers in compartment $i$. Both simulations and experiments show maximal rectification for parameter values ($d_G = 20\,\mu$m, $\alpha \simeq 30^\circ$) that are close to the theoretically predicted optimum (Fig.~3D). Thus, although the minimal model neglects variability in swimming behavior and hydrodynamic effects,  it captures the main features of the experiments. 
\par

Generally, our numerical and  experimental results support the hypothesis that rectification of microorganisms in environments with broken reflection symmetry is a ``universal'' phenomenon~\cite{2008Reichhardt_PRL,2010Aranson_PNAS,2010DiLeonardo_PNAS,2002AstumianHanggi,2009Hanggi,2009Stark,2011Stark,2002Reimann}. The ratchets mimic  Maxwell's demon but there is no conflict with thermodynamics due to the non-equilibrium nature of  living systems.

\begin{figure}[b]
\centering
  \includegraphics[clip=,width=   \columnwidth ]{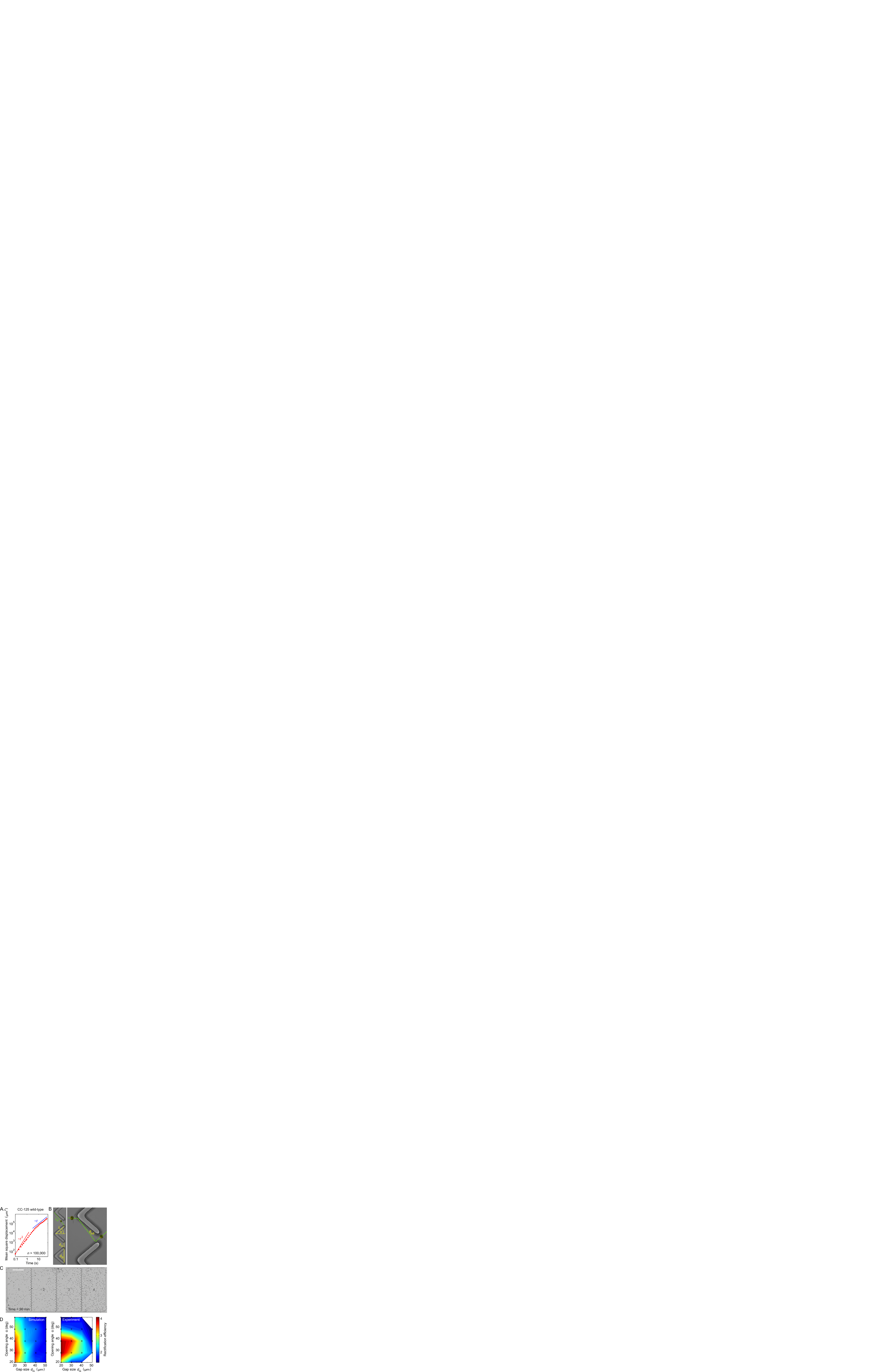}
\caption{
Rectification of algal locomotion in microfluidic ratchets. 
(A)~When confined in a quasi-2D chamber (height $25\,\mu$m), wild-type CC125 perform run-and-turn motions, moving ballistically (average speed $\bar{V}\sim30\,\mu$m/s) on short time scales ($< 2$ s) and diffusively on larger time scales.
(B)~Ratchet geometry and schematic representation of secondary algal scattering.
(C)~Rectified steady-state for a chamber with four compartments (Movie~S3). Scale bar 0.5 mm. 
(D)~In both simulations of the minimal model (Materials and Methods) and experiments,  the rectification efficiency $R=\langle N_4 \rangle/\langle N_1 \rangle$ exhibits a maximum near to the theoretically predicted optimal ratchet parameters. Each data point (circles) represents an average  over 3 to 5 different experiments. Rectification efficiencies are  linearly interpolated. The standard deviation for different experiments is less  than $30\%$ of the mean value.
}
\end{figure}

\section{Conclusions}

Our analysis of individual sperm-surface and alga-surface interactions shows that, in contrast to prevailing theoretical views, multiple flagellar contact determines the surface scattering of these eukaryotic cells. 
The ``puller''-swimmer \textit{Chlamydomonas reinhardtii} exhibits  a complex ciliary contact dynamics over the course of the scattering process, which spans a large number of successive swimming strokes and leads to an almost complete erasure of memory about the incidence angle. Direct contact of the flagella with nearby  walls may also be the main factor in the surface interactions of other motile algae, as the general structure of cilia is highly conserved across many eukaryotic species.  The flagella-induced self-trapping of bull spermatozoa and  the \textit{Chlamydomonas} mutant mbo1 suggests that  similar steric mechanisms could also be responsible for the long surface-residence times of other ``pusher''-type microorganisms~\cite{2009Tang_PRL,2011Drescher_PNAS}, possibly even in the case of bacteria, where the role of hydrodynamic interactions as the main determinant of surface scattering has recently been questioned~\cite{2009Tang_PRL,2011Tang_PRE,2011Drescher_PNAS}. With regard to future theoretical  studies, our  experimental results anticipate that a detailed qualitative and quantitative understanding of microbial surface interactions will require models that account for the elastic properties of eukaryotic cilia and bacterial flagella.

\par
Finally, as an illustration of how empirically measured surface-scattering laws can be exploited to control locomotion of unicellular green algae in biotechnological applications, we have demonstrated robust rectification of random swimming for wild-type \textit{Chlamydomonas} algae. In contrast to microfluidic rectification devices for rod-like prokaryotic (pusher) swimmers~\cite{2007AustinChaikin,2010Austin_PRL}, the optimal ratchet geometry for algae (pullers) exploits secondary scattering. The proposed design can be serialized (see Materials and Methods) and parallelized to facilitate large-scale microfluidic implementation.  Combining the  results of the scattering analysis and the subsequent rectification study suggests the possibility of integrating different ratchet geometries to create selection devices that sort microorganisms according to structure and dynamics of their propulsive appendages. More generally, the present investigation implies that suitably designed microstructured surfaces can yield new diagnostic tools to quantify function and response of eukaryotic cilia (mechano-sensing, desynchronisation after contact, etc.),  which may help to improve our understanding of transport processes in the respiratory or reproductive systems of higher organisms.  

\begin{acknowledgments}
The authors would like to thank Howard Berg and Paul Chaikin for invaluable discussions, and they are grateful to Elgin Columbo for providing the spermatozoa.
\end{acknowledgments}

\begin{materials}

\subsection{Sperm sample preparation} 
Cryogenically frozen bull spermatozoa were purchased from Genus Breeding LTD, UK. For each experiment, a sample of 250$\mu$L was thawed in a water bath (37$^\circ$C)  for 15s. The sample was  washed three times by centrifuging at 500$g$ for 5min and resuspending the pellet in a basic medium containing 72mM KCl, 160mM sucrose, 2mM Na-pyruvate and 2mM Na-phosphate buffer at pH 7.4~\cite{1984Rikmenspoel}.

\subsection{Algal growth} 
\textit{Chlamydomonas reinhardtii} strains CC-125 wild-type, CC-2347 shf1-277,  CC-2289 lf3-2,  and CC-2679 mbo1 (The Chlamydomonas Resource Center, \texttt{http://www.chlamy.org}) were grown axenically in Tris-Acetate-Phosphate (TAP) medium~\cite{2009Harris}, on an orbital shaker (200 rpm) in a diurnal growth chamber (KBW400, Binder). Daily cycle:  14h at 100$\mu$Em$^{-2}$s$^{-1}$ PAR (Fluora, OSRAM), 10h in dark, at 24$^\circ$C. Cells were harvested during the exponential growth phase. To achieve large concentrations of highly motile algae  ($\geq10^6$ cells/cm$^3$), cultures were centrifuged in Eppendorf tubes at 100$g$  for 10min and, subsequently, fast-swimmming algae were selected according to their ability to swim upwards against gravity. Thereafter, fresh TAP was added to the cell pellet, and the tubes were placed in the diurnal chamber for at least one hour to allow cells to recover.

\subsection{Microfluidics}
Quasi-2D channels were manufactured using standard soft lithography techniques~\cite{1998Whitesides}. The master mould was produced from SU8 2015 (MicroChem Corp.), spun to a 25$\mu$m thick layer and exposed to UV light through a high resolution mask to obtain the desired structures. The microfluidic chip containing the channels was cast from PDMS (Sylgard 184, Dow Corning) and bonded to a PDMS covered glass slide after oxygen plasma treatment of the surfaces. 
Microchannels for the sperm experiments were designed such that the sperm cells could be injected in a region that was spatially separated from the observation area. In the observation area, microchannels (100$\mu$m wide) were arranged in a zig-zag pattern with 90$^\circ$ corners, where the scattering events were imaged (Fig. 1A,B). The temperature was measured by a calibrated thermistor, which was inserted into the PDMS chip 2mm away from the observation region. To prevent adhesion of the spermatozoa to the walls of the channel, BSA at 5mg/mL was added prior the injection into the chamber. After injection of the sperm sample into the microchannel, motile spermatozoa could escape from the injection site into the observation region. The concentration of motile spermatozoa in this region did not exceed 1\% volume fraction. 
In both spermatozoa and \textit{Chlamydomonas} experiments, the microbial solutions were introduced through inlets that were plugged with unpolymerized PDMS afterwards. This procedure prevents fluid flows through the chambers and ensures conservation of the total number of cells over the course of the experiment.  In the  \textit{Chlamydomonas} experiments, the concentration of the algae was kept below 2\% volume fraction. For the rectification studies, each channel was subdivided into four chambers of size 2mm$\times$1mm$\times$0.025mm, separated by wedge-shaped barriers (Fig.~3). We treated PDMS surfaces of the channels prior the experiments with 10\% (w/v) Polyethylene glycol (m.w. 8000, Sigma) solution in water for 30min to prevent adhesion of the algae to the walls, and then flushed them gently with fresh TAP.  For the given parameters and uniform initial concentration profile across each chamber, rectified steady states were achieved after typically 90min (in the dark).

\subsection{Microscopy}
To identify the swimming characteristics of individual spermatozoa and  \textit{Chlamydomonas}, and their scattering distributions  (Fig.~1,2), cell trajectories  were reconstructed by applying a custom-made particle-tracking-velocimetry algorithm to image data taken with a Nikon TE2000-U inverted microscope (10x objective, 10fps). The flagella dynamics close to the boundary (Fig. 1A,B and~\ref{f:chlamy_scattering}A,B; Movies S1,S2 and S3, S4, S5 S6) was captured with a Fastcam SA-3 Photron camera (500-2000fps, 40x/NA 1.3 oil immersion and 60x/NA 1.0 water immersion objectives). For sperm white light was used and for  \textit{Chlamydomonas}  bright field illumination under red light ($\lambda>$620nm) to minimize phototaxis.
Sperm-surface scattering angles $\theta$ were determined by tracking the cell body up to a distance of 70$\mu$m from the corner. For  \textit{Chlamydomonas},  incidence and scattering angles $\theta_\text{in}$ and $\theta_\text{out}$ were obtained by measuring the slope of the trajectory at a distance of 20$\mu$m in either direction from the  scattering point (defined as the location at which the distance from the wall becomes minimal).  In the rectification experiments, algae concentrations in the microfluidic chambers were measured by averaging intensity profiles of image data obtained with a confocal scanning microscope Zeiss LSM 700 (5x objective). Here, the transmitted light PMT mode was used, while exposing the chambers to laser light (639nm) at the lowest intensity to minimize phototactic response of the algae.

\subsection{Numerical Simulations} 
We simulated the dynamics of  $N=750$  self-propelled, non-interacting point particles in a 2D box using MATLAB. Box size and ratchet geometries  where chosen to match the experimental set-up (Fig.~3). 
To account for the finite radius of \textit{Chlamydomonas}, a virtual layer of thickness $a=5\,\mu$m was added to boundaries and obstacles. Particles move ballistically at a constant speed $V$ until undergoing random turns, or colliding with boundaries or obstacles. Initial particle speeds $V$ were sampled from a Gamma-distribution $\Gamma(x;k,\nu)$ with parameters $k=4.2,\nu=7.3\mu$m/s, obtained from a best-fit to the experimentally measured speed distribution. Run-times between successive random turns are sampled from independent exponential distributions with mean~$\tau$. New directions after turn events are sampled uniformly from the unit circle. We simulated two types of boundary collision scenarios as simplified approximations to the experimentally observed wild-type scattering behavior:  deterministic collisions with an outgoing angle $\theta_\text{out}=16^\circ$, and randomized forward scattering. In the latter case,  the outgoing angles $\theta_\text{out}$ were sampled from a  truncated superposition of three Gaussians distributions $\Phi(x)=\sum_{i=1}^3 \lambda_i\Phi(x;\mu_i,\sigma_i)$ while rejecting angles  $\theta_\text{out}$ outside of the interval $(0,\pi)$. The distributions parameters were chosen as $(\lambda_1,\lambda_2)=(0.48,0.26)$ with normalization requiring $\lambda_3=1-\lambda_1-\lambda_2$, and $(\mu_1,\mu_2,\mu_3)=(0.43,0.26,3.06)$ and $(\sigma_1,\sigma_2,\sigma_3)= 
(0.58,0.14,0.55)$, obtained from a best-fit to the wild-type scattering distribution. In the deterministic case, our simulations confirmed maximal rectification for $\alpha\sim 37^\circ$. Simulation results in Fig.~3D show averages from three runs with~$\tau=1.5$~s for randomized scattering.  Generally, numerical results were found to be qualitatively robust under moderate variations of the model parameters, but the optimal wedge angle is sensitive to changes in the scattering distribution.

\subsection{Data Resources and Supplementary Movie Information} 
A MATLAB script of the source code,  simulation data and experimental raw data, and additional experimental movies can be downloaded from: 
\vspace{0.2cm}\\
http://damtp.cam.ac.uk/user/gold/datarequests.html
\vspace{0.2cm}\\
Scale bars  are 20$\mu$m in Movies~S1 and S2; 10$\mu$m in Movies S3, S4, S5, and S6; 0.5mm in Movie~S7. 

\subsection{Serialization (Markov Model)}
At low-to-intermediate algal volume fractions, as realized in our experiments, 
the dynamics of the \textit{Chlamydomonas}  population on the microfluidic chip can be 
described by a Markov model~\cite{2007AustinChaikin}. This allows to estimate the total rectification with  an increasing number of compartments. 
Assuming that the chip consists of $ i=1,\ldots, N$ identical compartments (Fig. 2D), the time evolution of the relative concentration $p_i(t)$ of algae in the $i$th compartment
is governed by
\begin{eqnarray}
\dot p_1 &=& - k_R\, p_1 + k_L\, p_2, 
\notag\\
\dot p_i &=&  k_R\, p_{i-1}- (k_L+k_R)\, p_i + k_L \,p_{i+1}, 
\\
\dot p_N &=&  k_R\, p_{N-1} -k_L\, p_{N}, \qquad\qquad 2\le i\le N-1
\notag,
\end{eqnarray}
where $k_{L/R}$ denote the rates for transitions to the left/right neighboring compartment ($\dot p_i\equiv dp_i/dt$ and $\sum_i p_i=1$). The stationary distribution $\{p^s_i\}$, corresponding to the eigenvector of the eigenvalue $\lambda=0$  of the transition matrix $K_{ij}$, defined by $\dot p_i=\sum_jK_{ij}p_j$,  is obtained as 
\begin{eqnarray}\label{e:stationary}
p^s_i=\frac{k_L^{N-i} k_R^{i-1}}{Z},
\qquad 
Z= \sum_{i}k_L^{N-i} k_R^{i-1}. 
\end{eqnarray}
The effective rates  $k_{L/R}$ can be estimated by fitting $p^s_i$ to the experimentally measured stationary distribution. Assuming $k_R>k_L$, Eq.~\eqref{e:stationary} implies that rectification increases with the number of chambers $N$ as $p^s_N/p^s_1=(k_R/k_L)^{N-1}$. Assuming detailed balance and diffusive backward flux suggests that $k_R/k_L=p_R (d_G+d_B)/d_G$, where $p_R$ is probability that an alga is  guided through the gap after having entered the barrier region. From our experimental data, we estimate $p_R\sim 0.3$.

\end{materials}

\end{article}

\end{document}